\documentclass{PoS}

\usepackage{epsfig}

\PoS{PoS(LAT2005)180}

\newcommand{\la}[1]{\label{#1}}
\newcommand{\ba}{\begin{eqnarray}}
\newcommand{\ea}{\end{eqnarray}}
\newcommand{\fig}{Fig.~}
\newcommand{\eq}{Eq.~}

\newcommand{\eqs}{Eqs.~}
\newcommand{\nr}[1]{(\ref{#1})}
\newcommand{\tr}{{\rm Tr\,}}
\renewcommand{\(}{\left(}
\renewcommand{\)}{\right)}
\newcommand{\lb}{\left\{}
\newcommand{\rb}{\right\}}

\def\lsi{\raise0.3ex\hbox{$<$\kern-0.75em\raise-1.1ex\hbox{$\sim$}}}
\def\gsi{\raise0.3ex\hbox{$>$\kern-0.75em\raise-1.1ex\hbox{$\sim$}}}

\newcommand{\msbar}{{\overline{\mbox{\rm MS}}}}
\newcommand{\tinyMSbar}{{\overline{\mbox{\tiny\rm{MS}}}}}
\newcommand{\lMS}{{\Lambda_\tinyMSbar}}
\newcommand{\Nf}{N_{\rm f}}
\newcommand{\Nc}{N_{\rm c}}
\newcommand{\Tc}{T_{\rm c}}
\newcommand{\bmu}{\bar\mu}
\newcommand{\sigs}{{\sigma_s}}
\newcommand{\rmi}[1]{{\mbox{\scriptsize #1}}}
\newcommand{\gE}{{g_\rmi{E}}}
\newcommand{\ggE}{{g_\rmi{E}^2}}

\newcommand{\mE}{{m_\rmi{E}}}
\newcommand{\mmE}{{m_\rmi{E}^2}}

\newcommand{\ggM}{{g_\rmi{M}^2}}

\title{Spatial string tension revisited}

\ShortTitle{Spatial string tension revisited}

\author{\speaker{York Schr\"oder}\\
        Faculty of Physics, University of Bielefeld, 
        D-33501 Bielefeld, Germany\\
        E-mail: \email{yorks@physik.uni-bielefeld.de}}

\author{Mikko Laine\\
        Faculty of Physics, University of Bielefeld, 
        D-33501 Bielefeld, Germany\\
        E-mail: \email{laine@physik.uni-bielefeld.de}}

\abstract{
The spatial string tension, a classic non-perturbative probe for the
convergence of the weak-coupling expansion at high temperatures,
can be determined in full QCD as well as in a dimensionally reduced
effective theory. Comparing both approaches, we find surprisingly
good agreement almost down to the critical temperature of the
deconfinement phase transition.\\[3cm]\mbox{}\hfill BI-TP 2005/37}

\FullConference{XXIIIrd International Symposium on Lattice Field Theory\\
		25-30 July 2005\\
		Trinity College, Dublin, Ireland}

\begin{document}

%%%%%%%%%%%%%%%%%%%%%%%%%%%%%%%%%%%%%%%%%%%%%%%%%%%%%%%%%%%%%%%%%%%%%%%%%%%

\section{Introduction}
\la{se:intro}

The interest in QCD at temperatures $T$ larger than (a few) hundred MeV
is triggered not only by purely theoretical reasons, but also
by ongoing heavy ion collision experiments, and by cosmology.
Given asymptotic freedom, a weak coupling expansion of this 
high-temperature phase seems well within reach. 
In practice, however, this expansion converges only slowly, 
and even shows a non-trivial analytic structure in the gauge 
coupling $g^2$.

By now, the problematic degrees of freedom have been identified. 
They are {\em soft} gauge-field modes with typical momenta $p\sim gT$,
which give rise to odd powers in $g$,
as well as {\em ultrasoft} modes $p\sim g^2T$, which 
enter the series via non-perturbative coefficients.
For parametrically small values of the coupling $g$,
these scales are well separated, such that an effective field theory
treatment becomes feasible.

The general picture is that perturbation theory should work fine 
for parametrically {\em hard} scales $p\sim 2\pi T$,  
while soft and ultrasoft scales need improved analytic schemes, 
or non-perturbative treatment.
We will work within dimensionally reduced effective theories,
in order to treat these different physical contributions separately,
in a consistent scheme with controllable errors.

It appears mandatory to give quantitative evidence for
the general picture sketched above.
To this end, the strategy is to pick some simple observables
and compare, as a function of $T$, {\em full} results 
(e.g. from 4d lattice QCD simulations~\cite{Boyd:1996bx})
with {\em predictions} from the soft/ultrasoft effective theory setup,
which should be exact for asymptotically large temperatures.
This has been done for e.g. static correlation lengths \cite{mu},
and in general agreement was found down to $T\sim 2\Tc$,
where $\Tc$ is the deconfinement phase transition temperature.

As another concrete example of an observable allowing for an 
unambiguous comparison, 
we discuss the spatial string tension $\sigs$ in this paper.
It is defined in a manifestly gauge invariant way as the coefficient
in the area law of a large rectangular Wilson loop
$W_s(R_1,R_2)$ in $(x_1,x_2)$ plane,
\ba\la{eq:1}
 \sigs \equiv - \lim_{R_1 \to \infty} \lim_{R_2 \to \infty} 
 \frac{1}{R_1 R_2} \ln W_s(R_1,R_2) \;.
\ea
It has been measured in SU(3) on the 4d lattice, 
as a function of the temperature $T$ 
(e.g. Ref.~\cite{Boyd:1996bx}),
\ba\la{eq:2}
 \frac{\sqrt{\sigs}}{T} &=& \phi_a\( \frac{T}{\Tc} \)\;.
\ea

Our aim here is to get the effective theory prediction for $\sigs$,
and to compare it with the lattice data, in order to assess the performance
of the effective theory setup \cite{Laine:2005ai}. 
In the following two sections, we sketch the 2-step perturbative 
matching process of 4d QCD onto 3d M{\footnotesize(agnetostatic)}QCD, 
and discuss convergence properties. 
In section \ref{se:results}, we take existing data on $\sigs$ from 
3d lattice MQCD, match it to 4d QCD, and compare with the 4d lattice data.

%%%%%%%%%%%%%%%%%%%%%%%%%%%%%%%%%%%%%%%%%%%%%%%%%%%%%%%%%%%%%%%%%%%%%%%%%%%

\section{Effective theory setup: QCD $\rightarrow$ EQCD}
\la{se:QE}

At high temperatures, all QCD dynamics is contained in a 
simpler, three-dimensional effective field theory called EQCD,
\ba\la{eq:3}
 {\cal L}_\rmi{E} &=& \frac12 \tr F_{kl}^2 + \tr [D_k,A_0]^2 + 
 \mmE \tr A_0^2 +\lambda_\rmi{E}^{(1)} (\tr A_0^2)^2
 +\lambda_\rmi{E}^{(2)} \tr A_0^4 + ... \;,
\ea
where $F_{kl}=i[D_k,D_l]/\gE$, $D_k=\partial_k-i\gE A_k$ 
with the dimensionful 3d gauge coupling $\gE$,
and the dots represent higher-order operators.
In order to correctly describe all contributions from hard and soft scales, 
the parameters of 3d EQCD have to be regarded as matching coefficients,
and are therefore related to the parameters of full QCD 
(being $g^2$, $T$, $\Nc$, $\Nf$, $\mu_\rmi{q}$, $m_\rmi{q}$).
Perturbative matching \cite{bn} gives, schematically,
\ba
 \mmE &=& T^2 \lb \#\, g^2+ \#\, g^4+...\rb\;,\la{eq:4a}\\ 
 \lambda_\rmi{E}^{(1),(2)} &=& T\lb \#\, g^4+\#\, g^6+...\rb\;,\la{eq:4b}\\
 \ggE &=& T\lb g^2+\#\, g^4+\#\, g^6+...\rb\;,\la{eq:4c}
\ea 
where all coefficients symbolized by ``$\#$'' above are known. 
Most can be conveniently read from e.g. Ref.~\cite{adjoint}, 
while the $g^6$ term in the last line has been obtained
only recently \cite{Laine:2005ai}.
Higher-order coefficients could be obtained straightforwardly
from the next order in the loop expansion.

There are also higher-order operators 
\cite{sc} in EQCD which become important at some point. 
In general, their relative magnitude can be estimated as~\cite{adjoint} 
\ba\la{eq:5}
 \delta {\cal L}_\rmi{E} \sim 
 g^2 \frac{D_k D_l}{(2 \pi T)^2} {\cal L}_\rmi{E} 
 \sim 
 g^2 \frac{(g^2T)^2}{(2 \pi T)^2} {\cal L}_\rmi{E}  
 \;,
\ea
where we assumed to be considering an observable 
dominated by the ultrasoft scale $p\sim g^2T$.
Thus, the relative magnitude is at most $\sim g^6$, 
smaller than any known terms in \eqs\nr{eq:4a}--\nr{eq:4c}.

At this point, having the first few terms of the perturbative
series of, say, $\ggE=\ggE(g^2,T)$ at hand, one may ask about 
its convergence properties. 
In practice, renormalization is needed of course: let
$g^2=g^2(\bmu)$ be the (4d QCD) $\msbar$ coupling. From 
the solution of the 2-loop renormalization group equation, 
we define the $\msbar$ scale parameter as usual, and
find the 2-loop running coupling as a function of $\bmu/\lMS$,
\ba
 \lMS &\equiv& \lim_{\bmu\to\infty}
 \bmu \Bigl[ b_0 g^2(\bmu) \Bigr] ^{-b_1/2 b_0^2}
 \exp \Bigl[ -\frac{1}{2 b_0 g^2(\bmu)}\Bigr] \;,\la{eq:6a}\\
 \frac{1}{g^2(\bmu)} &\approx& 2 b_0 \ln\frac{\bmu}{\lMS}
 + \frac{b_1}{b_0} \ln\biggl( 2 \ln\frac{\bmu}{\lMS} \biggr) \;,\la{eq:6b}
\ea
where $b_0 \equiv -\beta_0/2(4\pi)^2$, $b_1 \equiv -\beta_1/2(4\pi)^4$ 
are coefficients of the QCD beta function.
Hence, we can now write $\ggE=\ggE\(\bmu,\lMS,T\)=
T\,\phi_b\(\bmu/T,T/\lMS\)$ 
as a function of two dimensionless variables.

Formally, the renormalization scale dependence is of higher order,
while numerically, there is $\bmu$ dependence due to our truncation
of the perturbative series. 
We are free to choose some optimization procedure, e.g. the 
{\em principle of minimal sensitivity}, according to which we
choose $\bmu_\rmi{opt}$ as the extremum of the 1-loop expression 
for $\ggE$.
This leaves us $\ggE=T\,\phi_c\(T/\lMS\)$ as a function of
one variable only, which is plotted in the left panel of 
\fig\ref{fig:gT} for $\Nf=3$.
Comparing 1-loop and 2-loop expressions (the gray band shows the effect
of a scale variation within $\bmu=(0.5 ... 2.0) \times \bmu_\rmi{opt}$), 
note that the process of perturbative matching shows very comforting 
convergence properties: corrections are in the 10-20\% range, and scale
dependence gets significantly reduced.

%%%%%%%%%%%%%%%%%%%%%%%%%%%%%%%%% FIGURE %%%%%%%%%%%%%%%%%%%%%%%%%%%%%%%%%
\begin{figure}[t]
\centerline{%
\epsfxsize=7cm\epsfbox{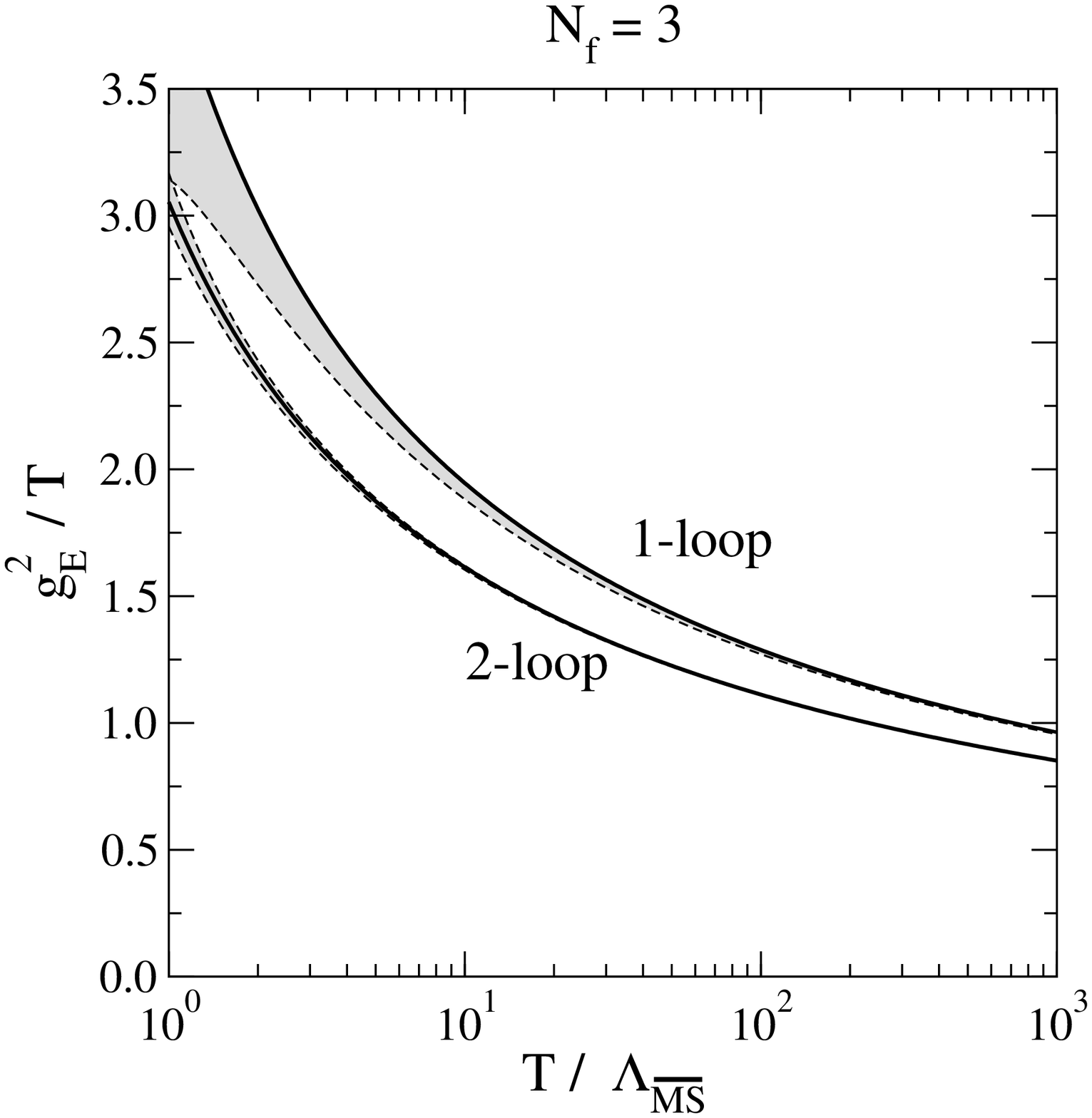}%
\epsfxsize=8cm\epsfbox{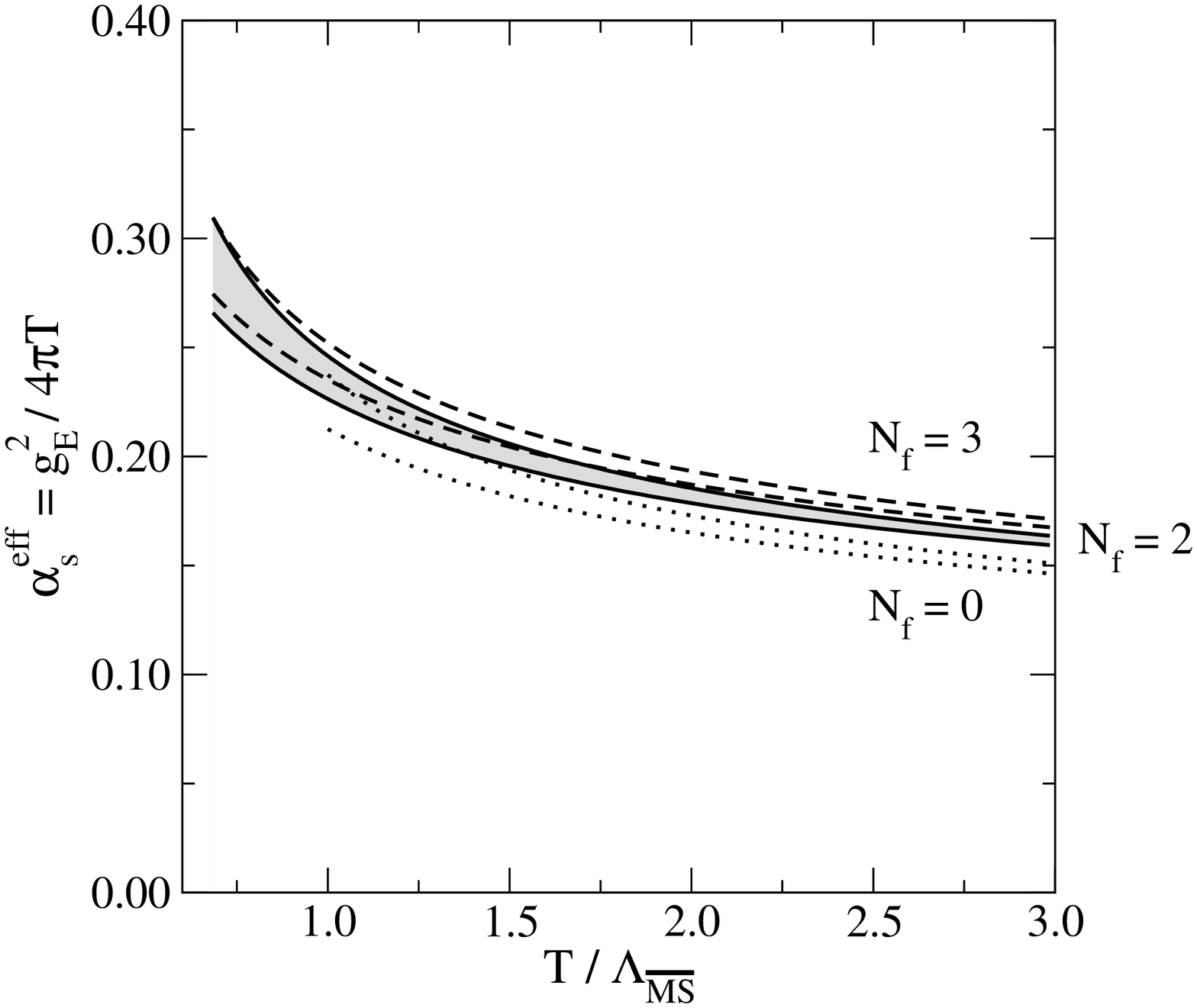}}
\caption[a]{Left panel: {\it The 1- and 2-loop values for $\ggE/T$. 
For each $T$, the scale $\bmu$ has been fixed to $\bmu_\rmi{opt}$ 
as explained in the text. The gray band corresponds to a variation
of $\bmu_\rmi{opt}$ within a factor of two.} 
Right panel: {\it The effective 2-loop gauge coupling of EQCD,
$\alpha_\rmi{s}^\rmi{eff}\equiv \ggE/4\pi T$,
for various values of $\Nf$. Scale dependence results from $\bmu_\rmi{opt}$
variation as before. Note the different ranges of temperatures on
the horizontal axes.}}
\la{fig:gT}
\end{figure}
%%%%%%%%%%%%%%%%%%%%%%%%%%%%%%%%%%%%%%%%%%%%%%%%%%%%%%%%%%%%%%%%%%%%%%%%%%%

In the right panel of \fig\ref{fig:gT}, we show the effective gauge coupling 
$\alpha_\rmi{s}^\rmi{eff}\equiv \ggE/4\pi T$ of EQCD,
for several $\Nf$, in a much smaller temperature interval close to the 
phase transition temperature $\Tc\sim\lMS$. 
Noting that this 3d effective coupling is reasonably small even at these 
low temperatures, we are led yet again to observe that treating the hard
modes perturbatively appears well justified. 

%%%%%%%%%%%%%%%%%%%%%%%%%%%%%%%%%%%%%%%%%%%%%%%%%%%%%%%%%%%%%%%%%%%%%%%%%%%

\section{Effective theory setup: EQCD $\rightarrow$ MQCD}
\la{se:EM}

The low-energy behaviour of 3d EQCD is contained in another
three-dimensional effective field theory, called MQCD,
\ba\la{eq:7}
 {\cal L}_\rmi{M}  &=&  \frac12 \tr F_{kl}^2+... \;.
\ea
As before, the dots stand for higher-order operators, 
while the matching coefficients can be determined
perturbatively \cite{pg,Laine:2005ai}
\ba\la{eq:8}
 \ggM &=& \ggE\lb 1+\#\, \frac{\ggE}{\mE}
 +\#\, \frac{g_\rmi{E}^4}{\mmE} 
 +\#\, \frac{\ggE \lambda_\rmi{E}^{(1),(2)}}{\mmE} +...\rb \;.
\ea
Let us note here -- without showing the corresponding plot -- 
that this expansion converges extremely well, even close to $\Tc$.
Hence, we can safely ignore higher loop corrections for $\ggM$.

The higher-order operators of MQCD,
\ba\la{eq:9}
 \delta {\cal L}_\rmi{M} \sim \ggE \frac{D_k D_l}{m_\rmi{E}^3} 
 {\cal L}_\rmi{M} \sim 
 \ggE \frac{(g^2 T)^2}{m_\rmi{E}^3} {\cal L}_\rmi{M} 
 \;,
\ea
give a relative contribution parametrically smaller than 
any of the known terms in \eq\nr{eq:8}, 
and will be neglected in the following. 

%%%%%%%%%%%%%%%%%%%%%%%%%%%%%%%%%%%%%%%%%%%%%%%%%%%%%%%%%%%%%%%%%%%%%%%%%%%

\section{Results}
\la{se:results}

We are now in a position to write down the effective theory 
prediction for the spatial string tension $\sigs$, \eq\nr{eq:1}.
The observable $\sigs$ exists not only in 4d QCD, but also in 
3d SU(3) gauge theory, which is nothing but MQCD, \eq\nr{eq:7}.
Since the 3d gauge coupling is dimensionful, and furthermore is the
only scale that MQCD possesses, naive dimensional analysis dictates
$\sigs = \#\, g_\rmi{M}^4$. 
The proportionality constant is non-perturbative, and can be measured  
by 3d lattice simulations. Taking most recent lattice data \cite{mt}, 
\ba\la{eq:10}
 \frac{\sqrt{\sigs}}{\ggM} &=& 0.553(1) \;.
\ea 

%%%%%%%%%%%%%%%%%%%%%%%%%%%%%%%%% FIGURE %%%%%%%%%%%%%%%%%%%%%%%%%%%%%%%%%
\begin{figure}[t]
\centerline{\epsfysize=8.0cm\epsfbox{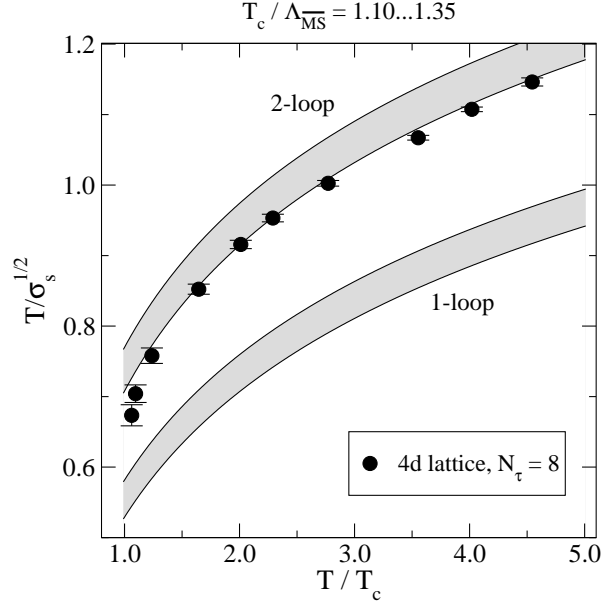}}
\caption[a]{\it Comparison of 4d lattice data for the spatial string 
tension \cite{Boyd:1996bx} with expressions obtained by combining 1-loop
and 2-loop results for $\ggE$ together with \eq\nr{eq:8} and
the non-perturbative
value of the string tension of 3d SU(3) gauge theory, \eq\nr{eq:10}.
The upper edges of the bands correspond to $\Tc/\lMS = 1.35$,
the lower edges to $\Tc/\lMS = 1.10$.} 
\la{fig:compare}
\end{figure}
%%%%%%%%%%%%%%%%%%%%%%%%%%%%%%%%%%%%%%%%%%%%%%%%%%%%%%%%%%%%%%%%%%%%%%%%%%%

To compare with the 4d lattice results of the form shown in 
\eq\nr{eq:2} (see \fig\ref{fig:compare}), 
we need to relate $\ggM$ and $T$.
First, using \eq\nr{eq:8} and \eqs\nr{eq:4a}--\nr{eq:4c}, 
\ba\la{eq:11}
 \frac{\sqrt{\sigs}}{T} 
 &=& 0.553(1)\; \frac{\ggM}{\ggE}\; \frac{\ggE}T 
 \;=\; \phi_d\(\frac{T}{\lMS}\) \;.
\ea
Next, we need to relate $\lMS$ and $\Tc$.
This is in fact a classic problem in (4d) lattice QCD.
One line of measurements~\cite{bb} employs the $T=0$ string tension 
to get~\cite{ltw} 
\ba\la{eq:12}
 \frac{\Tc}{\lMS}&=&\frac{\Tc/\sqrt{\sigma}}{\lMS/\sqrt{\sigma}} 
 \;=\; 1.16(4) \;,
\ea
while another possibility is to go via the Sommer scale \cite{sn}
\ba\la{eq:13}
 \frac{\Tc}{\lMS}&=&\frac{r_0\Tc}{r_0\lMS} 
 \;=\; 1.25(10) \;.
\ea 
To be conservative, we will consider the interval 
$\Tc/\lMS=1.10 ...1.35$, which also incorporates the
result of Ref.~\cite{gupta}.

In \fig\ref{fig:compare}, we finally compare the 3d effective theory 
prediction for $\sigs$ (gray bands) with the 4d lattice data (black dots).
As a caveat, note that the lattice data has not been extrapolated
to the continuum limit.
On the other hand, we stress that the comparison is parameter-free. 
We may take the excellent agreement of the 2-loop prediction
with the lattice data as
support for hard/soft+ultrasoft picture of thermal QCD.

To conclude, 
we have given yet another example of a static observable in thermal
QCD, for which the program of dimensional reduction works well,
even down to temperatures $T\sim 2\Tc$.

%%%%%%%%%%%%%%%%%%%%%%%%%%%%%%%%%%%%%%%%%%%%%%%%%%%%%%%%%%%%%%%%%%%%%%%%%%%

\end{document}